\documentclass[amsmath,amssymb,superscriptaddress,nobalancelastpage,prl,twocolumn]{revtex4-1}

\usepackage{bm}
\usepackage{braket}
\usepackage{diagbox}
\usepackage{graphicx}
\usepackage{placeins}
\usepackage{siunitx}
\usepackage{ulem}
\usepackage{varioref}
\usepackage{hyperref}
\usepackage[all]{hypcap}
\hypersetup{colorlinks,linkcolor=blue,urlcolor=blue,citecolor=blue}

\newcommand{\LCOchem}{La$_2$CuO$_4$}
\newcommand{\LSCOchem}{La$_{1.77}$Sr$_{0.23}$CuO$_4$}

\newcommand{\EuLSCOchem}{La$_{1.59}$Eu$_{0.2}$Sr$_{0.21}$CuO$_4$}

\newcommand{\Tlchem}{Tl$_{2}$Ba$_{2}$CuO$_{6+\delta}$}
\newcommand{\Bichem}{Bi$_{1.74}$Pb$_{0.38}$Sr$_{1.88}$CuO$_{6+\delta}$}
\newcommand{\Prchem}{Pr$_{1.15}$La$_{0.7}$Ce$_{0.15}$CuO$_{4}$}
\newcommand{\polsig}{$\bar{\sigma}$}
\newcommand{\polpi}{$\bar{\pi}$}
\newcommand{\tg}{$t_{2g}$}
\newcommand{\dxy}{$d_{xy}$}
\newcommand{\dxz}{$d_{xz}$}
\newcommand{\dyz}{$d_{yz}$}
\newcommand{\dz}{$d_{3z^2-r^2}$}
\newcommand{\dx}{$d_{x^2-y^2}$}
\newcommand{\daovera}{$d_{\mathrm{A}} / a$}

\begin{document}

 \author{K.~P.~Kramer}
     \affiliation{Physik-Institut, Universit\"{a}t Z\"{u}rich, Winterthurerstrasse 190, CH-8057 Z\"{u}rich, Switzerland}
     
 \author{M.~Horio}
   \affiliation{Physik-Institut, Universit\"{a}t Z\"{u}rich, Winterthurerstrasse 190, CH-8057 Z\"{u}rich, Switzerland}
   
    \author{S.~S.~Tsirkin}
  \affiliation{Physik-Institut, Universit\"{a}t Z\"{u}rich, Winterthurerstrasse 190, CH-8057 Z\"{u}rich, Switzerland}
   
    \author{Y.~Sassa}
  \affiliation{Department of Physics and Astronomy, Uppsala University, SE-75121 Uppsala, Sweden}
  
 \author{K.~Hauser}
   \affiliation{Physik-Institut, Universit\"{a}t Z\"{u}rich, Winterthurerstrasse 190, CH-8057 Z\"{u}rich, Switzerland}
   
\author{C.~E.~Matt}
   \affiliation{Physik-Institut, Universit\"{a}t Z\"{u}rich, Winterthurerstrasse 190, CH-8057 Z\"{u}rich, Switzerland}
   \affiliation{Swiss Light Source, Paul Scherrer Institut, CH-5232 Villigen PSI, Switzerland}
   
   \author{D.~Sutter}
     \affiliation{Physik-Institut, Universit\"{a}t Z\"{u}rich, Winterthurerstrasse 190, CH-8057 Z\"{u}rich, Switzerland}
     
 \author{A.~Chikina}
 \affiliation{Swiss Light Source, Paul Scherrer Institut, CH-5232 Villigen PSI, Switzerland}
 
\author{N.~Schr\"oter}
 \affiliation{Swiss Light Source, Paul Scherrer Institut, CH-5232 Villigen PSI, Switzerland}
 
\author{J.~A.~Krieger}
 \affiliation{Laboratory for Muon Spin Spectroscopy, Paul Scherrer Institute, CH-5232 Villigen PSI, Switzerland} \affiliation{Laboratorium f\"ur Festk\"orperphysik,  ETH Z\"urich, CH-8093 Z\"urich, Switzerland}
 
\author{T.~Schmitt}
 \affiliation{Swiss Light Source, Paul Scherrer Institut, CH-5232 Villigen PSI, Switzerland}
 
\author{V.~N.~Strocov}
 \affiliation{Swiss Light Source, Paul Scherrer Institut, CH-5232 Villigen PSI, Switzerland}
 
\author{N.~Plumb}
 \affiliation{Swiss Light Source, Paul Scherrer Institut, CH-5232 Villigen PSI, Switzerland}
 
\author{M.~Shi}
 \affiliation{Swiss Light Source, Paul Scherrer Institut, CH-5232 Villigen PSI, Switzerland}
 
\author{S.~Pyon}
\affiliation{Department of Advanced Materials, University of Tokyo, Kashiwa 277-8561, Japan}
\author{T.~Takayama}
\affiliation{Department of Advanced Materials, University of Tokyo, Kashiwa 277-8561, Japan}
\author{H.~Takagi}
\affiliation{Department of Advanced Materials, University of Tokyo, Kashiwa 277-8561, Japan}
   
  \author{T.~Adachi}
\affiliation{Department of Engineering and Applied Sciences, Sophia University, Tokyo 102-8554, Japan}

\author{T.~Ohgi}
\affiliation{Department of Applied Physics, Tohoku University, Sendai 980-8579, Japan}

\author{T.~Kawamata}
\affiliation{Department of Applied Physics, Tohoku University, Sendai 980-8579, Japan}

 \author{Y.~Koike}
 \affiliation{Department of Applied Physics, Tohoku University, Sendai 980-8579, Japan}
 
 \author{T.~Kondo}
 \affiliation{ISSP, University of Tokyo, Kashiwa, Chiba 277-8581, Japan}

\author{O.~J.~Lipscombe}
         \affiliation{H. H. Wills Physics Laboratory, University of Bristol, Bristol BS8 1TL, United Kingdom}
         
     \author{S.~M.~Hayden}
       \affiliation{H. H. Wills Physics Laboratory, University of Bristol, Bristol BS8 1TL, United Kingdom}
     
\author{M.~Ishikado}
       \affiliation{Comprehensive Research Organization for Science and Society (CROSS), Tokai, Ibaraki 319-1106, Japan}
       
\author{H.~Eisaki}
       \affiliation{Electronics and Photonics Research Institute, National Institute of Advanced Industrial Science and Technology, Ibaraki 305-8568, Japan}

\author{T.~Neupert}
  \affiliation{Physik-Institut, Universit\"{a}t Z\"{u}rich, Winterthurerstrasse 190, CH-8057 Z\"{u}rich, Switzerland}

\author{J.~Chang}
    \affiliation{Physik-Institut, Universit\"{a}t Z\"{u}rich, Winterthurerstrasse 190, CH-8057 Z\"{u}rich, Switzerland}

\title{Band Structure of Overdoped Cuprate Superconductors:  Density Functional Theory Matching Experiments}
    
\begin{abstract}
A comprehensive angle resolved photoemission spectroscopy study of the band structure in single layer cuprates is presented with the aim of uncovering universal trends across different materials.
Five different hole- and electron-doped cuprate superconductors (\EuLSCOchem, \LSCOchem, \Bichem, \Tlchem, and \Prchem) have been studied with special focus on the bands with predominately $d$-orbital character.
Using light polarization analysis, the $e_g$ and $t_{2g}$ bands are identified across these materials.
A clear correlation between the \dz\ band energy and the apical oxygen distance $d_\textup{A}$ is demonstrated. 
Moreover, the compound dependence of the \dx\ band bottom and the $t_{2g}$ band top is revealed.
Direct comparison to density functional theory (DFT) calculations employing hybrid exchange-correlation functionals demonstrates excellent agreement.
We thus conclude that the DFT methodology can be used to describe the global band structure of overdoped single layer cuprates on both the hole and electron doped side.
\end{abstract}

\maketitle
  
\textit{Introduction:} 
The physics of cuprate superconductors has been a subject of intense investigations for more than three decades~\cite{LeeRMP2006,KeimerNat2015,ScalapinoRMP2012}. 
Yet, some of the most fundamental questions related to high-temperature superconductivity remain open.
For example, consensus on the mechanism underpinning cuprate superconductivity is still missing. 
Related to this is the question of the defining parameters for the transition temperature $T_\mathrm{c}$~\cite{JASlezakPNAS2008,NormanPNAS2008,HirofumiPRL10,MattNatCommun2018,WeberEPL2012,Adler2018}, and how to optimize it.
Starting point for most theoretical approaches to superconductivity is an (effective) electronic band structure as well as the interactions that are relevant for driving a pairing mechanism.
The former is typically obtained through density functional theory (DFT).
However, because DFT cannot describe all relevant aspects of the electronic structure (such as the Mott insulating phase out of which superconductivity emerges upon hole or electron doping~\cite{FurnessNatCom2018}) it is commonly viewed as too simplistic of an approach in the context of the cuprates~\cite{HozoiSREP11}.
Another widespread assumption is that effective models for cuprates can be constructed solely on the \dx\ band structure. 
This latter assumption has recently been challenged~\cite{HirofumiPRL10,HirofumiPRB12} by angle-resolved photoemission spectroscopy (ARPES) observations of a second band (\dz) hybridizing with the \dx\ orbital in 
overdoped \LSCOchem\ (LSCO)~\cite{MattNatCommun2018,Horio2018Dirac}.

Here we provide a systematic ARPES and DFT study of the electronic $d$-band structure across single layer cuprate superconductors. 
Five different hole- and electron-overdoped superconducting systems [\EuLSCOchem\ (Eu-LSCO), \LSCOchem\ (LSCO), \Bichem\ (Bi2201), \Tlchem\ (Tl2201), and \Prchem\ (PLCCO)] 
have been investigated experimentally.
This has led to three main observations:
(\textit{i}) clear identification of the \dz\ band position in three of the mentioned compounds,
(\textit{ii}) compound dependence of the \dx\ band bottom positions and
(\textit{iii}) the $t_{2g}$ (\dxy, \dxz\ and \dyz) band positions at the zone corner.
These experimental observations are quantified as a function of apical oxygen distance $d_\textup{A}$ and compared directly to DFT calculations. 
Generally, excellent quantitative agreement between DFT and experimental band structures is found.
It is therefore concluded that even though DFT is not capturing low-energy self-energy effects, it is successfully describing the global band structure of the cuprates.

\begin{figure*}
 	\begin{center}
 		\includegraphics[width=0.85\textwidth]{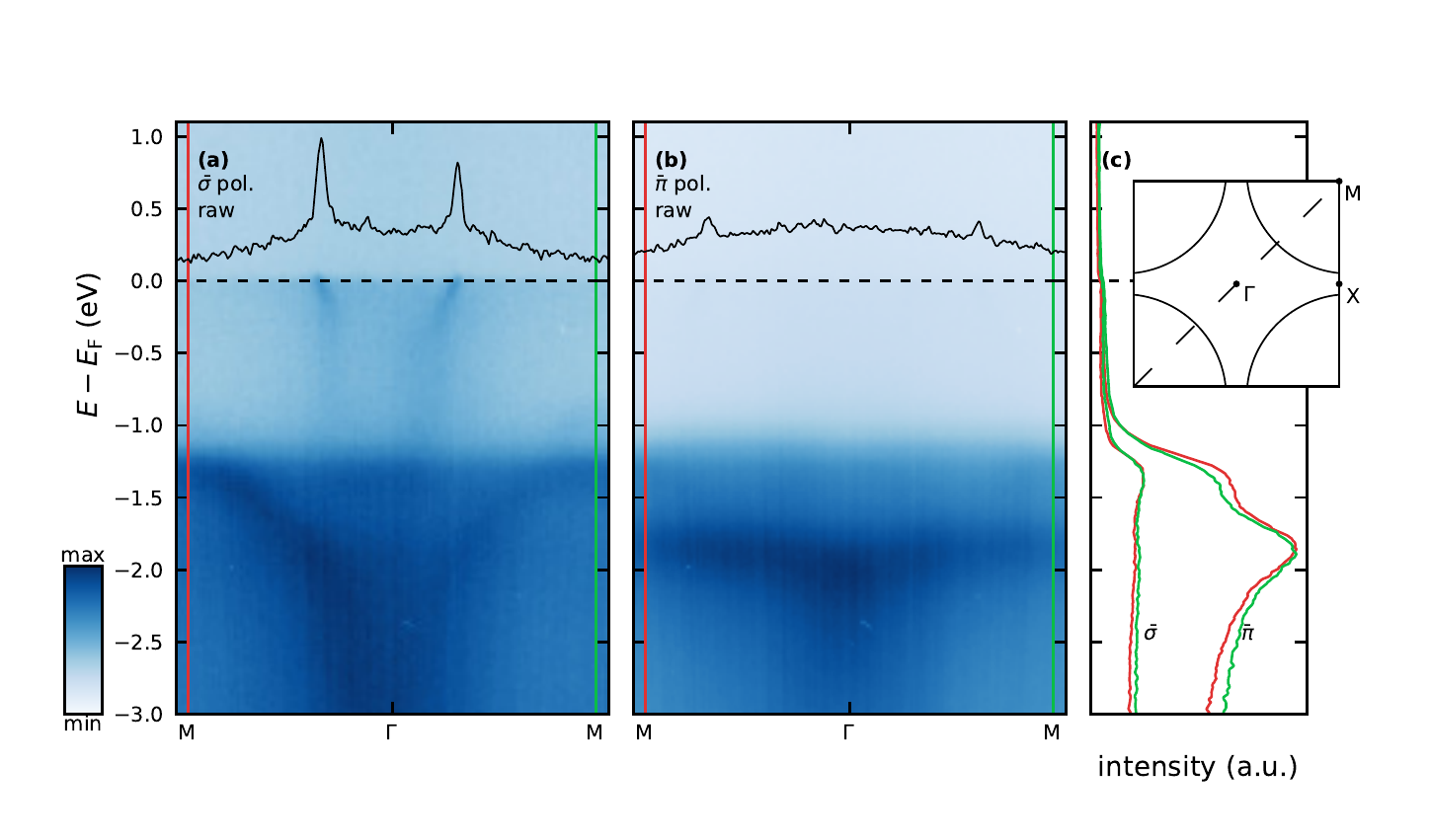}
 	\end{center}
 	\caption{Nodal soft x-ray ARPES spectra recorded on \Bichem\ with \SI{420}{\eV} incident photons of different linear polarizations, as indicated.
 	\textbf{(a),(b)} Recorded ARPES spectra along the line $\overline{\mathrm{M}\Gamma\mathrm{M}}$ (dashed line in the inset of (c)) for \polsig- and \polpi-polarized light, respectively.
 	Depicted at the top are momentum distribution curves taken at the Fermi level $E_\textup{F}$.
 	\textbf{(c)} Energy distribution curves taken along vertical lines of corresponding colors in (a) and (b). Inset: schematic of the Fermi surface with the diagonal (nodal) direction indicated as a dashed line.
 	}

	\label{fig:bi2201_raw}
 \end{figure*}
 
\textit{Methods:}
Single crystals of Eu-LSCO, LSCO, Bi2201,  Tl2201 and PLCCO were grown by floating zone or flux growth techniques.
Both ultraviolet (\SIrange{20}{200}{\eV}) and soft x-ray (\SIrange{200}{600}{\eV}) ARPES experiments were carried out at the SIS~\cite{SIS} and ADRESS~\cite{StrocovJSR2014} beamlines at the Swiss Light Source. 
All data were recorded at a temperature of approximately \SI{20}{K}.
Electrons were analyzed through a slit oriented within the photoemission mirror plane~\cite{DamascelliRMP03}.
Light polarization perpendicular (parallel) to the mirror plane is denoted as \polsig\ (\polpi).

Predicting the correct energies for the electronic bands is a notorious problem for DFT in many materials, which is mostly due to the unknown form of the exchange-correlation functional. 
Improvements over local density approximations are commonly obtained using hybrid functionals which mix in a portion $\alpha$ of exact exchange from Hartree-Fock theory~\cite{HSE06}. 
While $\alpha$ is a free parameter in general, we find good agreement between the theoretical and experimental band structures for all compounds studied by fixing $\alpha=0.1$. 
We thus propose this value as generically suited for cuprate superconductors.
A hypothetical tetragonal structure of \LCOchem\ with lattice parameters corresponding to overdoped LSCO was used and the chemical potential adjusted to match the actual hole-filling.
Similarly, for Bi2201, Tl2201 and PLCCO stochiometric tetragonal crystal structures were used as a starting point for the DFT calculations.
We ensured, on the example of Bi2201, that using an orthorhombic crystal structure leads to essentially the same results after downfolding the calculated band structure to the tetragonal Brillouin zone.
More details on the methodology used can be found in the supplementary material~\footnote{The supplementary material can be found at the end of this document. References 31-39 appear therein.}.
Although some of the systems studied have orthorhombic structures, we represent all data in tetragonal notation~\cite{DamascelliRMP03}, using the CuO$_2$ plaquette Brillouin zone nomenclature.
Therefore M and X, respectively, denote the zone corner $(1,1)$ and  boundary $(1,0)$ in units of $\pi/a$ with $a$ being the tetragonal in-plane lattice parameter [Fig.~\ref{fig:bi2201_raw} (c), inset].

\begin{figure*}
 	\begin{center}
 		\includegraphics[width=0.99\textwidth]{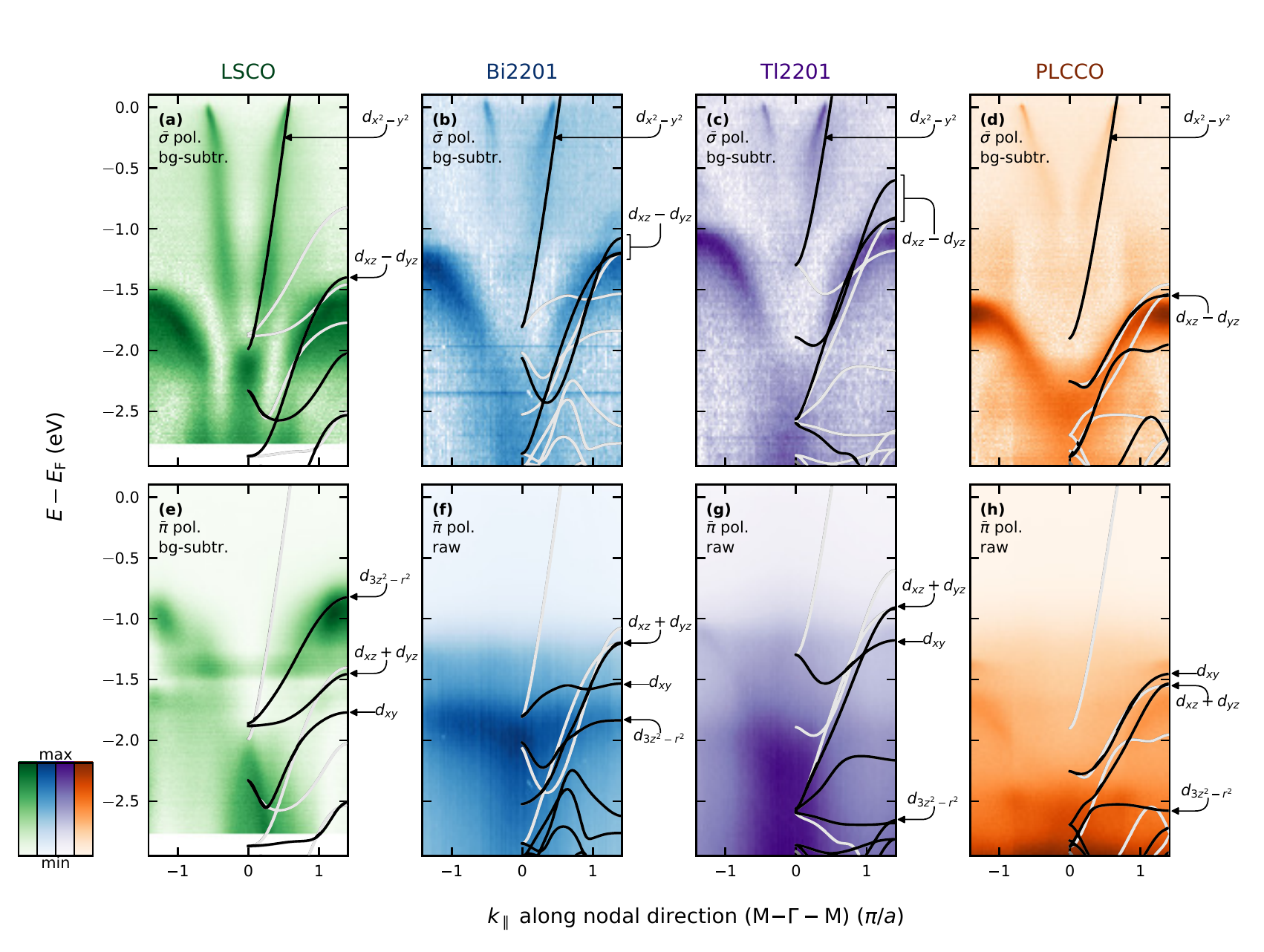}
 	\end{center}
 	\caption{Nodal ARPES spectra and DFT band structure calculations for the single-layer cuprates \LSCOchem, \Bichem, \Tlchem\ and \Prchem.
 	\textbf{(a)}--\textbf{(d)} Background subtracted nodal ARPES spectra recorded using \polsig-polarized light.
 	\textbf{(e)}--\textbf{(h)} Nodal spectra recorded with \polpi-polarized light.
 	The spectrum in (e) has received the same background subtraction as the ones in (a) -- (d). 
    By contrast, the spectra in panels (f) -- (h) represent raw data as the background subtraction methodology is not advised in presence of flat bands (see text and discussion in suppl. Fig.~\ref{fig:bg_subtraction}).
 	The calculated DFT band structure is overlaid with solid lines where black and light grey colors indicate finite or zero matrix elements, respectively. 
    The labeling of the orbital character is only valid at the M point for the cases of \dxz, \dxy\ and \dyz. See suppl. Fig.~\ref{fig:orbital_characters} for a full orbital character assignment.
 	}
 	
	\label{fig:arpes_vs_dft}
 \end{figure*}

\textit{Results:} 
We start by examining the nodal spectra recorded on overdoped Bi2201 using linearly polarized soft x-rays, see Fig.~\ref{fig:bi2201_raw}. 
The photoemission intensities of the observed band structure are highly dependent on the incident light polarization. Three distinct bands are identified.
(\textit{i}) The intensely-studied nodal quasiparticle dispersion~\cite{ZhouNat2003,FatuzzoPRB2014,YoshidaJCMP07} crossing the Fermi level $E_{\textup{F}}$, which is observed with $\bar{\sigma}$-polarized light only. 
This fact can be appreciated both from the energy distribution maps (EDMs) and the momentum distribution curves (MDCs) at the Fermi level [Figs.~\ref{fig:bi2201_raw}~(a),(b)].
(\textit{ii}) A second dispersive band with energy maximum of \SI{-1.3}{eV} at the M-point and band bottom at the $\Gamma$-point. 
At the M-point,  this band is featured in both the $\bar{\sigma}$ and $\bar{\pi}$ channels [Fig.~\ref{fig:bi2201_raw}~(c)].
Notice that in comparison to previous studies of Bi2201, our data displays extremely weak $(\pi,\pi)$-folded replica bands~\cite{KingPRL2011,RosenNatComm2013}.
As a result and in contrast to Refs.~\onlinecite{XiaPRL2008,MeevasanaPRB2007}, this dispersive band is not found at the $(\pi,\pi)$-folded equivalent $\Gamma$-point.
(\textit{iii}) The \polpi-channel features an additional weakly dispersive band at energy \SI{-1.8}{eV} [Fig.~\ref{fig:bi2201_raw}~(b)].
This band, which to the best of our knowledge has not been reported previously, is completely suppressed in the \polsig-sector.

\begin{figure}
 	\begin{center}
 		\includegraphics[width=0.49\textwidth]{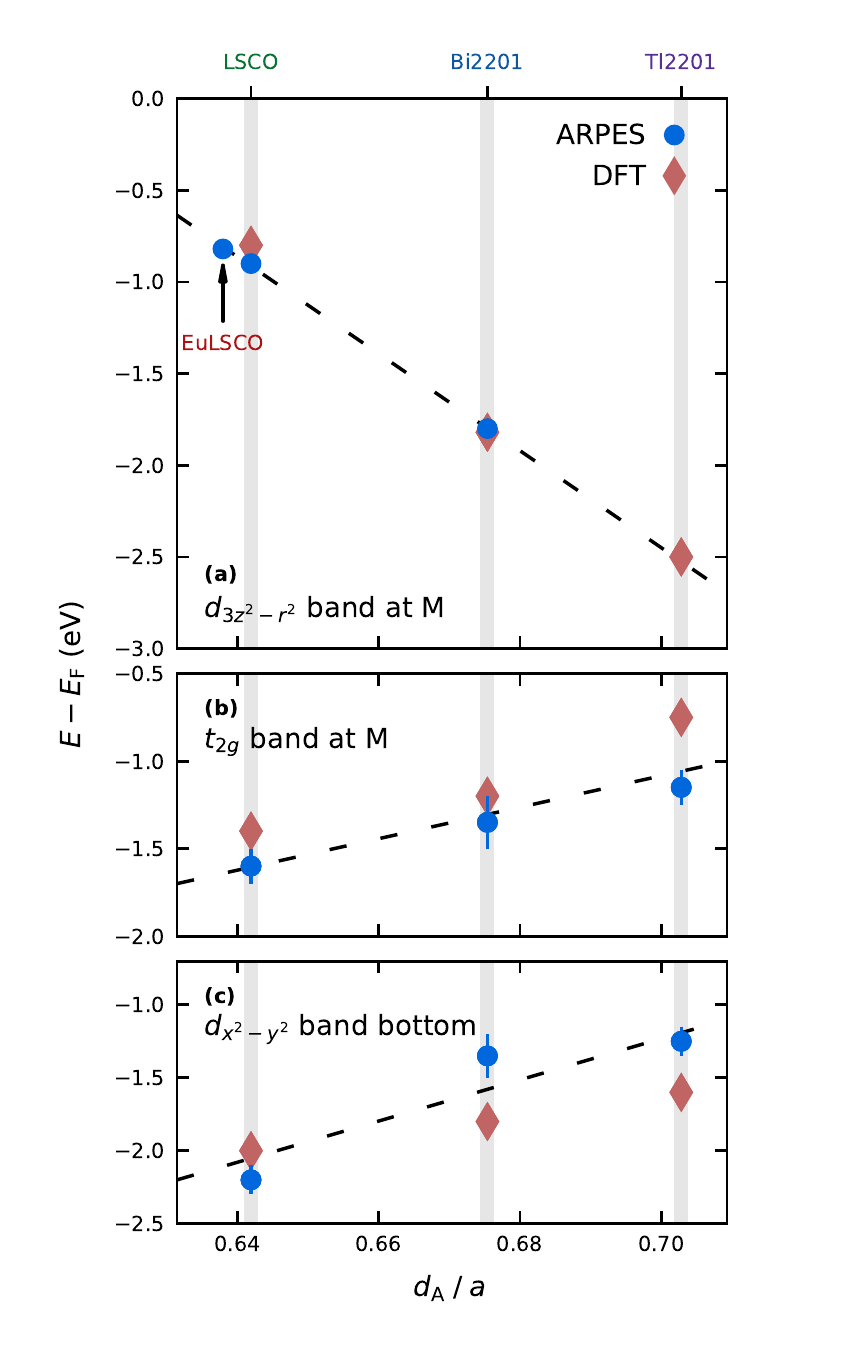}
 	\end{center}
 	\caption{Band structure characteristics versus ratio between apical oxygen distance and (tetragonal) in-plane lattice constant \daovera. Blue points represent the ARPES experiments whereas results from our DFT calculations are indicated by red diamonds. 
 	Dashed lines represent guides to the eye.
 	\textbf{(a)} \dz\ band position at the M point versus \daovera.
 	\textbf{(b)} Position of the \tg\ band at the zone corner M versus \daovera.
 	Mean values of the \dxy, \dxz+\dyz\ and \dxz-\dyz\ band positions at M are taken for the DFT points.
 	\textbf{(c)} \dx\ band bottom as function of \daovera.}
	\label{fig:results}
 \end{figure}

Nodal ARPES spectra recorded in $\bar{\sigma}$ and $\bar{\pi}$ polarization on LSCO, Bi2201, Tl2201 and PLCCO are shown in Figs.~\ref{fig:arpes_vs_dft}~(a)--(h). 
For all four compounds, the band crossing the Fermi level is visible (suppressed) in the $\bar{\sigma}$ ($\bar{\pi}$) channel. 
Interestingly, the bottom of this band varies significantly, from \SI{-2}{eV} in LSCO to \SI{-1.25}{eV} for Tl2201.
In the $\bar{\pi}$ channel an additional band feature appears for LSCO, Bi2201 and PLCCO. 
The position and band width of this $\bar{\sigma}$-suppressed band differs for the three compounds. 
In LSCO, it disperses from \SI{-0.9}{eV} at the M-point to \SI{-1.5}{eV} at the zone center, while for Bi2201 and PLCCO the $\bar{\pi}$-branches found at \SI{-1.8}{eV} and \SI{-2.5}{eV} respectively, are quasi non-dispersive.
Finally, for Tl2201 no band unique to the $\bar{\pi}$ channel was identified down to \SI{-3}{eV}.
 
The DFT band structure of LSCO, Bi2201, Tl2201 and PLCCO [Figs.~\ref{fig:arpes_vs_dft}~(a)--(h)] has been calculated as described above and in the supplementary material. 
In addition to the band dispersions, their expected photoemission matrix elements are indicated according to their mirror eigenvalues.
To first order, photoemission matrix elements can be understood through simple symmetry considerations \cite{DamascelliRMP03,moser17experimentalist}.
Our experimental setup has a mirror-plane defined by the incident photon beam and the electron analyzer.
With respect to this plane, the electromagnetic field $\bm{A}$ has even (odd) parity for parallel (perpendicular) \polpi\ (\polsig) polarization [suppl. Figs.~\ref{fig:structures}~(a), (b)].
Meanwhile, the photoemitted electron wave function has even parity.
The mirror eigenvalues of the (\dx, \dz, \dxy, \dxz+\dyz, \dxz$-$\dyz) orbital states are ($-1$, 1, 1, 1, $-1$), respectively (suppl. Tab.~\ref{table:table1}).
Therefore, \dz\ and \dxy\ states will be suppressed in the \polsig\ channel while \dx\ states cannot appear under \polpi\ illumination.
Since \dxz\ and \dyz\ orbitals are not eigenstates of the mirror operator, electronic states along $\overline{\Gamma\mathrm{M}}$ are formed by even (\dxz\ + \dyz) and odd (\dxz\ $-$\dyz) linear combinations and will thus be visible under both polarizations.
In Fig.~\ref{fig:arpes_vs_dft}, bands with matrix element 0 are colored light grey, while bands with nonzero matrix element remain black.

For LSCO, the \dz\ band width is roughly \SI{0.5}{eV} with a band maximum at the M-point (\SI{-0.8}{eV}).
The crossing of the \dz\ and \dx\ bands, constituting a type-II Dirac line node, is protected by mirror symmetry~\cite{Horio2018Dirac,TaoPRB2018}. 
For Bi2201 and Tl2201, the \dz\ band is pushed to lower energies and an overall smaller \dz\ band width is found.
As a result, the nodal crossing of the \dz\ and \dx\ bands is not
found for Bi2201 and Tl2201. The type-II Dirac line node is thus unique to LSCO~\cite{Horio2018Dirac}.
We also notice that in LSCO, the \dz\ band lies above the $t_{2g}$ bands whereas the opposite is true for Bi2201 and Tl2201.
Finally, in comparison to LSCO, the bottom of the \dx\ band is closer to the Fermi level in Tl2201.

\textit{Discussion:}
The polarization dependence of the band which crosses the Fermi level allows us to assign it uniquely to the \dx\ orbital in all studied compounds.
For Bi2201, the next band below the Fermi level is found in both the $\bar{\sigma}$  and $\bar{\pi}$ channels and hence can be assigned to the \dxz, \dyz\ orbitals. 
The flat band found around \SI{-1.8}{eV} in the $\bar{\pi}$ channel has to have either \dxy\ or \dz\ character. 
For a unique orbital assignment we stress the following facts:
(\textit{i}) Resonant inelastic x-ray scattering (RIXS) measurements of the $dd$-excitations found the \dz\ states at \SI{-2}{eV} and below the \dxy, \dxz, \dyz\ states~\cite{PengNatPhys2016}. 
(\textit{ii}) As the \dxy\ orbital extends purely in the $xy$-plane, the \dxy\ band is generally expected to disperse strongly along the nodal direction. 
This is indeed confirmed by our DFT calculations.
Combined, this lets us assign the \SI{-1.8}{eV} band in Bi2201 to the \dz\ orbital.
As previously discussed in Ref.~\onlinecite{MattNatCommun2018} and shown in Fig.~\ref{fig:arpes_vs_dft} and suppl. Fig.~\ref{fig:bg_subtraction} the \dz\ band is clearly identified in La-based cuprates.
In the case of Tl2201, by contrast, no evidence  for the \dz\ band is found down to \SI{-3}{eV}. 
Therefore, either the \dz\ band in Tl2201 is pushed to even lower binding energies or this band is too faint to be observed. The latter scenario is supported by the fact that in PLCCO with no apical oxygen, the \dz\ band is found at around \SI{-2.5}{eV} [see Fig.~\ref{fig:arpes_vs_dft}(h)].

We plot the observed \dz\ band position at the M-point as a function of the compound dependent ratio between apical oxygen distance $d_\mathrm{A}$ and in-plane lattice constant $a$ (Fig.~\ref{fig:results}).
Furthermore, the \dx\ band bottoms and the energies of the $t_{2g}$ bands at the zone corner M are plotted alongside their respective positions found from our DFT calculations. One can see that the calculations capture the most salient band structure trends:
(\textit{i}) DFT correctly predicts how the \dz\ band position -- with respect to the Fermi level -- evolves as a function of \daovera. 
(\textit{ii}) DFT yields the right trends for the band widths of both the \dz\ and \dx\ bands. 
The \dz\ band width is, for example, gradually reduced when going through the series LSCO $\rightarrow$ Bi2201 $\rightarrow$ Tl2201.
The ARPES data on LSCO and Bi2201 supports that trend. 
For the \dx\ band, ARPES only reveals the occupied part. 
Instead of band width, it thus makes more sense to consider the band bottom.
It turns out that the \dx\ band minimum is varying across LSCO, Bi2201, and Tl2201 and the positions agree between DFT and experiment.
(\textit{iii}) The $t_{2g}$ band position at the zone corner also follows the trend of moving closer to $E_\textup{F}$ with increasing \daovera, both in experiment and the DFT calculations.
However, we stress that because the DFT methodology is not including electron interactions, it is not capturing self-energy effects such as the much discussed nodal waterfall structure~\cite{GrafPRL2007,VallaPRL2007,ChangPRB2007}.

\textit{Conclusions:} 
In summary, we have carried out a comprehensive ARPES and DFT study of the band structure across single layer cuprate superconductors.
Experimentally, five different overdoped cuprate compounds were studied using light polarization analysis to assign band orbital characters.
Both the $e_g$ (\dx\ and \dz) and $t_{2g}$ bands were discussed and their band positions and band widths were compared to DFT calculations. 
The excellent agreement between DFT and experimental results led us to conclude that the DFT methodology with proper choice of exchange-correlation functional does capture the global electronic structure of the overoped cuprates.

\textit{Acknowledgments:} 
K.P.K., M.H., D.S., J.A.K., and J.C. acknowledge support by the Swiss National Science Foundation. Y.S. is funded by the Swedish Research Council (VR) with a Starting Grant (Dnr. 2017-05078) O.K.F. and M.M. are supported by a VR neutron project grant (BIFROST, Dnr. 2016-06955). 
ARPES measurements were performed at the ADRESS and SIS beamlines of the Swiss Light Source at the Paul Scherrer Institute.
We thank the beamline staff for their support.

\nocite{VASP1,VASP2,PAW1,PAW2,HSE06,PBE-1996,Mostofi2008,Mostofi2014,KaminskiPRB2004,MattNatCommun2018,Horio2018Dirac,horio18threedimensional}


\clearpage
\onecolumngrid
\section*{Supplementary Information}

\renewcommand{\thefigure}{S\arabic{figure}}
\setcounter{figure}{0}
\renewcommand{\thetable}{S\Roman{table}}
\setcounter{table}{0}
\renewcommand{\theHtable}{Supplement.\thetable}
\renewcommand{\theHfigure}{Supplement.\thefigure}

\begin{figure*}[hb]
    \centering
    \includegraphics{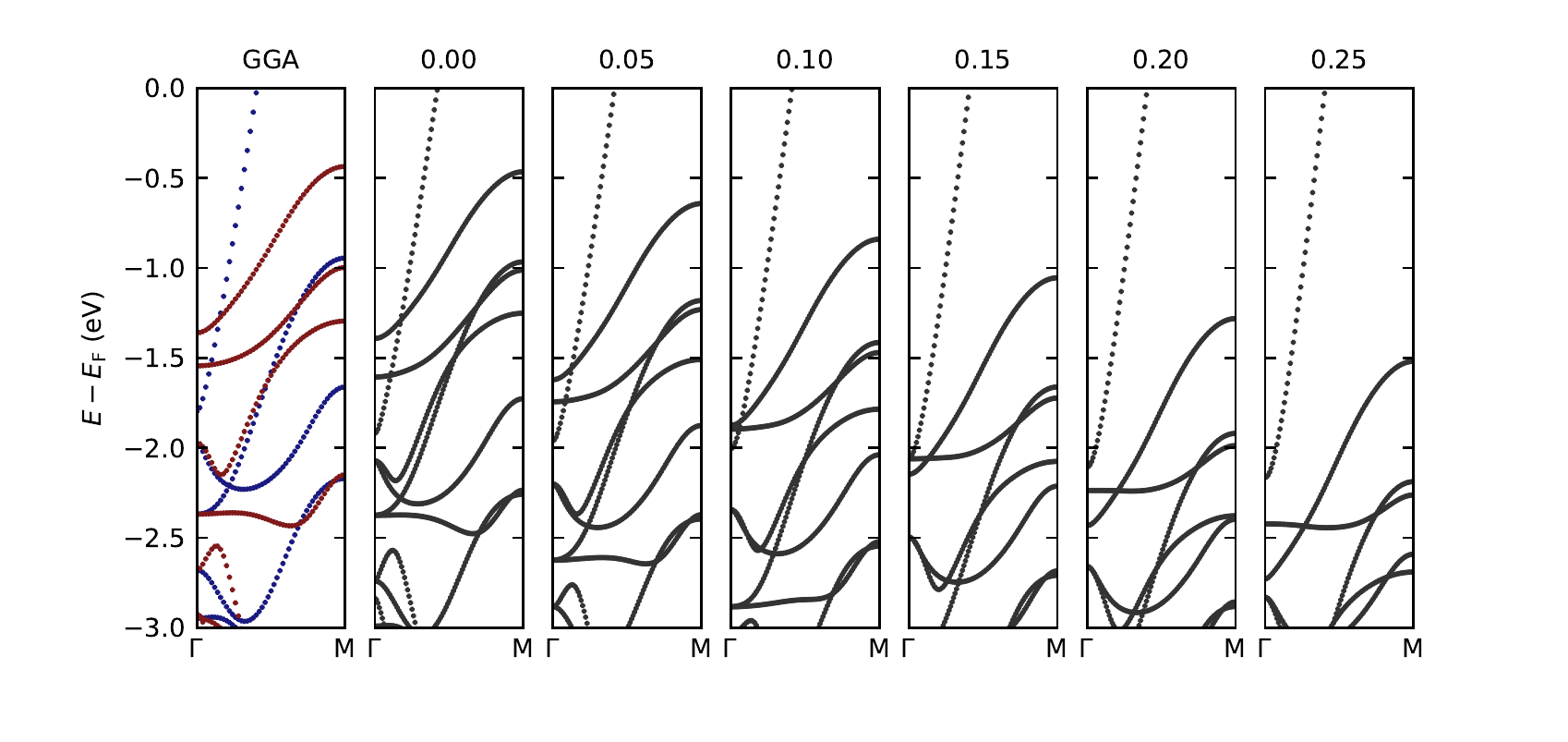}
    \caption{DFT results for different values of the amount of exact exchange $\alpha$ as well as pure GGA at the example of LSCO. Red and blue colors denote mirror eigenvalues of $+1$ or $-1$ respectively.}
    \label{fig:dft}
\end{figure*}

\textit{DFT calculations:}
The presented DFT calculations were performed using the {\tt VASP} code package~\cite{VASP1,VASP2} employing the projector-augmented wave method (PAW) \cite{PAW1,PAW2}. 
The exchange-correlation functional was treated in the form similar to the Heyd-Scuseria-Ernzerhof (HSE06) \cite{HSE06} screened hybrid functional, but with a variable portion $\alpha$ of Hartree-Fock (HF) exchange.
The standard HSE06 functional uses $\alpha=0.25$, i.e. 25\% of HF exchange and 75\% of Perdew, Burke, and Ernzerhof generalized-gradient approximation (GGA)~\cite{PBE-1996} for the short-range part of the exchange functional.
We discover that this tends to overestimate the binding energy of the bands in the materials under study.
On the other hand the pure GGA functional ($\alpha=0$) underestimates the binding energies.
By varying the portion of HF exchange, we find that the best overall agreement for all investigated materials is achieved by letting $\alpha=0.1$ (suppl. Fig.~\ref{fig:dft}).
After performing self-consistent hybrid-functional calculations on a regular 6x6x6 $\Gamma$-centered k-point grid, we interpolate the band structures along the $\overline{\Gamma\mathrm{M}}$ line by means of the {\tt Wannier90} code package~\cite{Mostofi2008,Mostofi2014}.

We add two notes concerning the calculations of PLCCO:
(\textit{i}) We started with the stochiometric crystal structure of Pr$_2$CuO$_4$ and simulated the La occupation by replacing one of the two Pr sites in the unit cell with La.
Even though this does not exactly represent the exact occupation that is present in the experimental compound, we ensured by comparison to calculations on Pr$_2$CuO$_4$ that varying the La content has only a minor effect on the electronic band structure.
Thus the small error in occupation should not affect our conclusion.
(\textit{ii}) The energies of the $4f$-bands resulting from our calculations lie above $E_\textup{F}$ due to the shifting of the Fermi level in order to account for the right electron filling.
This result is of course unphysical and we conclude that our methodology, while proving very successful for the $d$-orbitals, is unfit to correctly predict the $f$-orbitals.

\begin{figure*}[htb]
    \centering
    \includegraphics{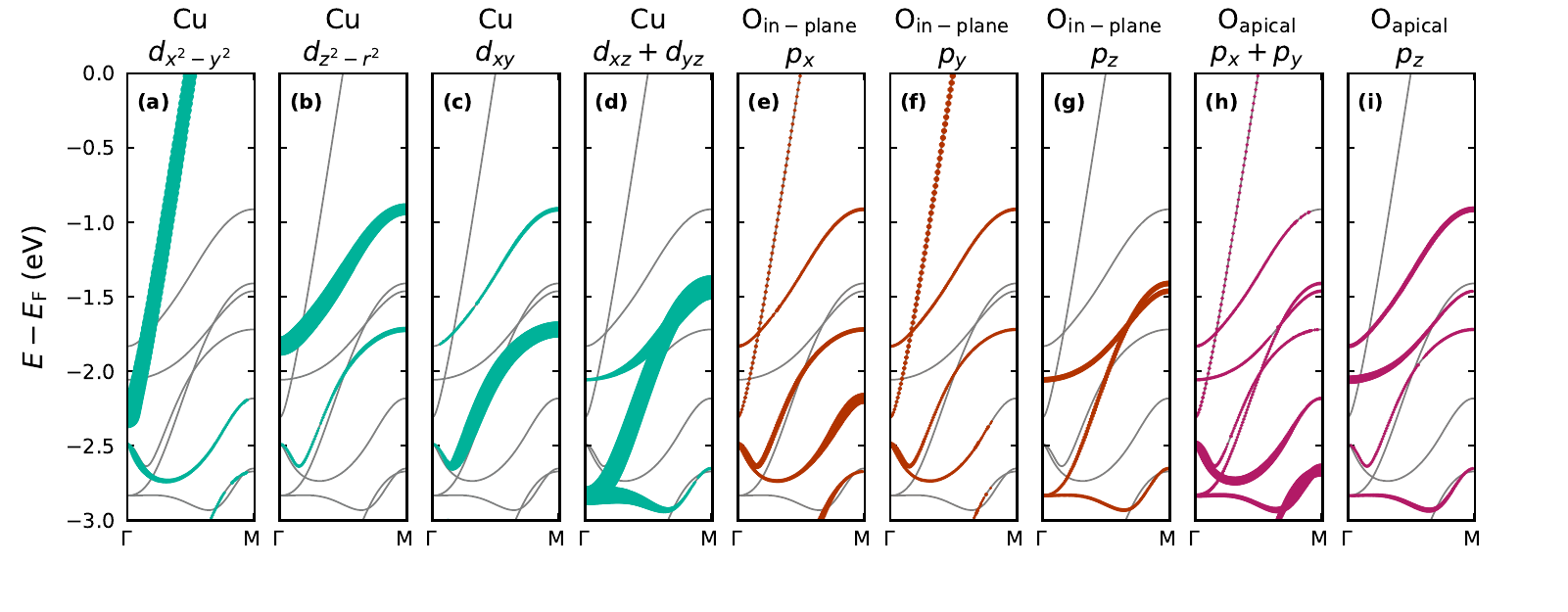}
    \caption{Orbital weights given by DFT at the example of LSCO for the Cu $d$-orbitals [panels (a)--(d)], the in-plane Oxygen $p$-orbitals [(e)--(g)] and the apical (out-of-plane) oxygen $p$-orbitals [(h), (i)].
    The marker area is proportional to the respective orbital weight.
    Other orbitals do not contribute significantly to the band structure in the shown energy range.}
    \label{fig:orbital_characters}
\end{figure*}


\textit{Background subtraction:}
In virtually all spectroscopy methods, the obtained data contains an intrinsic and extrinsic background.
For the cuprates, it has been demonstrated that these background contributions have significant dependencies on binding energy~\cite{KaminskiPRB2004}. 
In fact, the background at binding energies of \SI{1}{eV} might be an order of magnitude larger than at or near the Fermi level.
For comparison of band structures across a wide range of binding energy, it can therefore be useful to subtract the background intensity~\cite{MattNatCommun2018}. 
Assuming a $k$-independent background, 
we estimate the background profile by averaging the five lowest intensity points of each MDC~\cite{MattNatCommun2018,Horio2018Dirac}.
In this fashion, we are subtracting a constant background at each binding energy. MDCs therefore remain unchanged.
Two examples, using Eu-LSCO and Tl2201, of this background subtraction procedure are shown in supplementary Fig.~\ref{fig:bg_subtraction}. 
We stress that for flat non-dispersive bands such a background determination is not advisable as the band structure will be eliminated in the subtraction procedure.
We therefore only apply the background subtraction, in Fig.~\ref{fig:arpes_vs_dft} of the main text, to enhance the visibility of dispersive bands.

\begin{figure*}
 	\begin{center}
 		\includegraphics[width=1\textwidth]{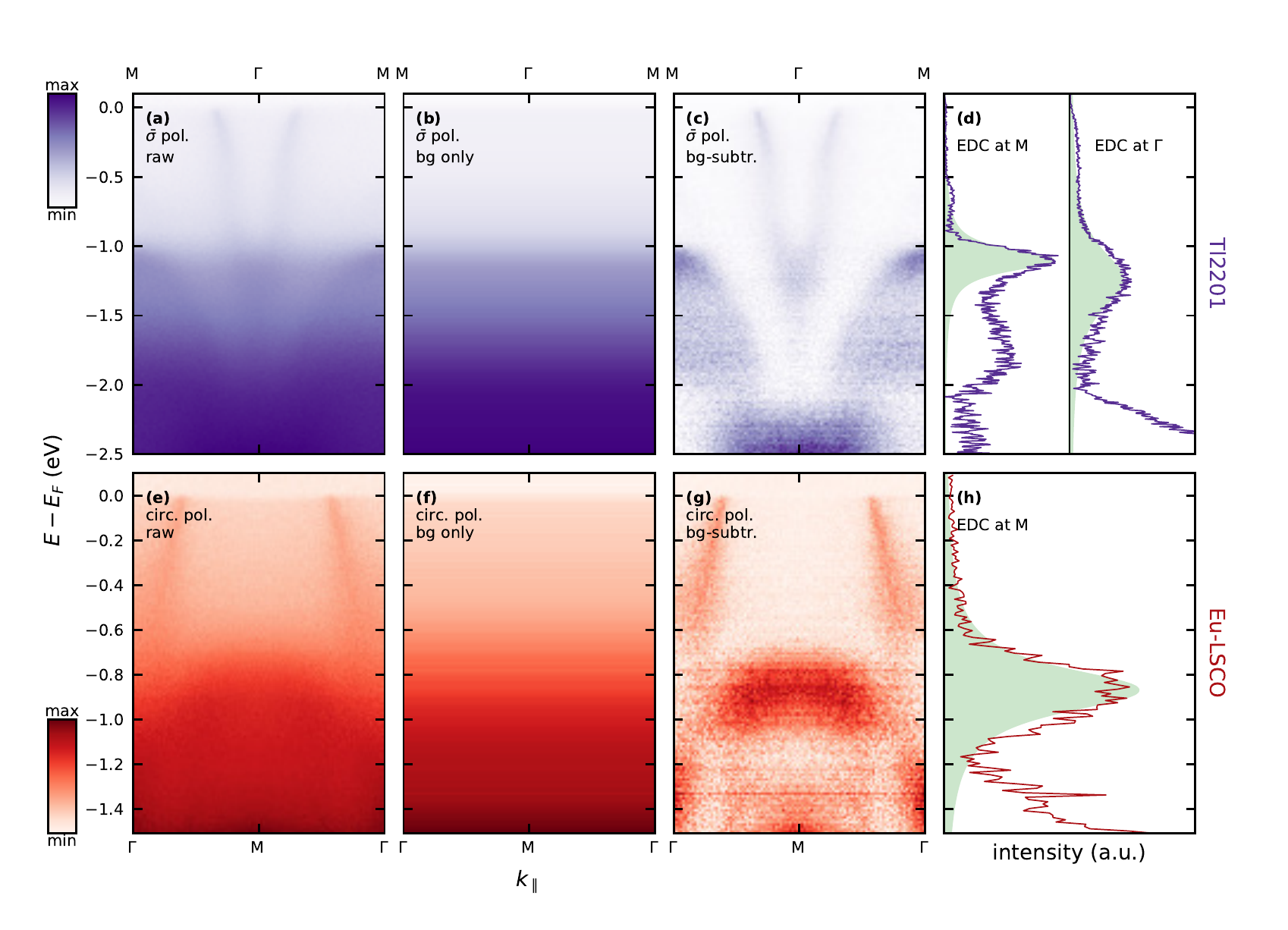}
 	\end{center}
 	\caption{Demonstration of  background subtraction, exemplified by  nodal spectra recorded on \Tlchem\ and \EuLSCOchem. 
 	\textbf{(a),(e)} Raw nodal spectra for Tl2201 and Eu-LSCO respectively.
 	\textbf{(b),(f)} Corresponding background profile extracted by averaging the lowest five points in every momentum distribution curve (see text).
 	\textbf{(c),(g)} Resulting background subtracted energy distribution maps.
 	\textbf{(d),(h)} Background subtracted energy distribution curves at the $\Gamma$- and/or M- point. 
 	In the case of Eu-LSCO, the data has been taken in the second Brillouin zone and was symmetrized around M.
 	A power-law mapping $y=x^\gamma$ with $\gamma=0.5$ has been applied to the colorscale in panels (a), (b), (e) and (f) in order to enhance visibility.
 	}
	 \label{fig:bg_subtraction}
 \end{figure*}
 
 \begin{figure*}
    \centering
    \includegraphics{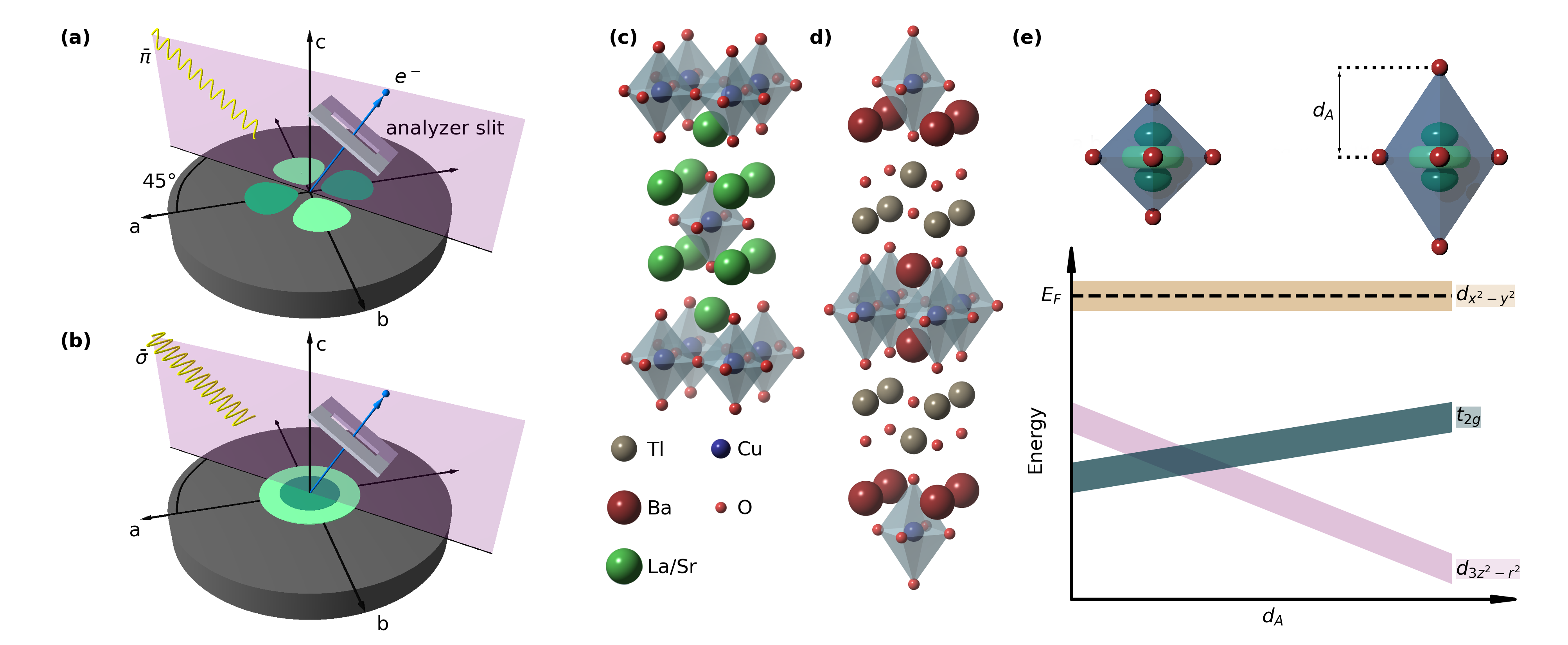}
    \caption{Photoemission geometry, crystal structures and crystal field splitting. 
    \textbf{(a),(b)} Schematic of the photoemission experimental setup. The projection of the Cu 3\dx/\dz\ orbital is visualized at the origin of the coordinate system and a nodal mirror plane defined by the incident photons and photoemitted electrons is indicated. Incoming light polarized perpendicular and parallel with respect to the mirror plane is labeled $\bar{\sigma}$ and $\bar{\pi}$, respectively.
    \textbf{(c),(d)} Crystal structures of LSCO and Tl2201, respectively. 
    \textbf{(e)} Top: Octahedra of different apical oxygen distance $d_\mathrm{A}$ with a representation of the \dz\ orbital enclosed. 
    Bottom: Schematics of the \dx, \dz\ and $t_{2g}$
    crystal field splittings as a function of apical oxygen distance $d_\mathrm{A}$. }
    \label{fig:structures}
\end{figure*}

\textit{Experimental details:} 
Eu-LSCO, LSCO and PLCCO data presented has been acquired at the SIS beamline with incident photon energies of \SI{160}{eV}, \SI{160}{eV} and \SI{55}{eV},  respectively.
Meanwhile, Bi2201 and Tl2201 data was taken at the ADRESS beamline with photon energies of \SI{420}{eV} and \SI{428}{eV}, respectively.
All data presented was acquired at a temperature of approximately \SI{20}{K}.
The used photon energies ensure that the data taken stems from planes through $\Gamma$ along the $k_z$-direction for LSCO, Tl2201 and PLCCO.
For Bi2201 we are closer to the Brillouin zone edge along the $k_z$-direction.
Meanwhile, all calculations were done for $k_z=0$, i.e. in planes through the Brillouin zone center.
This fact does not invalidate our comparison between ARPES and DFT, however:
The $k_z$-dispersion has been shown to be of very small (yet finite) order in LSCO~\cite{horio18threedimensional,MattNatCommun2018} and is not expected to be any bigger for Bi2201.
The effect of being at a different $k_z$ therefore only contributes a shift in the order of the marker size in Fig.~\ref{fig:results} of the main text.

\begin{table*}[h]
	\begin{center}
	{\tabcolsep = 3mm
	\begin{tabular}{rccccc}
		\hline \hline
		$\ket{d_j}$	   & $\ket{x^2-y^2}$ &  $\ket{3z^2-r^2}$ & $\ket{xy}$ & $\ket{xz}+\ket{yz}$ & $\ket{xz}-\ket{yz}$ \\   \hline
		$\bra{d_j} M_{xy} \ket{d_j}$ & $-1$ & $+1$ & $+1$ & $+1$ & $-1$ \\
		Character & odd       & even        & even & even    & odd    \\ 
        $\bra{f} A\bm{\epsilon}_{\bar{\sigma}} \ket{d_j}$ & $\neq$0 & 0 & 0 & 0 & $\neq$0 \\ 
	    $\bra{f} A\bm{\epsilon}_{\bar{\pi}} \ket{d_j}$ & 0 & $\neq$0 & $\neq$0 & $\neq$0 & 0 \\\hline
		\hline
	\end{tabular}}
	\end{center}
	\caption{Connection between the orbital mirror symmetries and photoemission matrix elements.
	The first row gives the eigenvalues with respect to a mirror at the nodal plane $M_{xy}$ for the $d$-orbitals in a basis in which $M_{xy}$ is diagonal.
	The last two rows contain the photoemission matrix element as a function of $d$-orbital and incident light polarization.
	$A$ denotes the magnitude of the electromagnetic vector potential.
	The polarization vector $\bm{\epsilon}_{\bar{\sigma}}$ ($\bm{\epsilon}_{\bar{\pi}}$) is perpendicular (parallel) to the analyzer plane and thus corresponds to $\bar{\sigma}$-polarized ($\bar{\pi}$-polarized) light of \textit{odd} (\textit{even}) character.
	The final state $\bra{f}$ is assumed to be a plane wave of \textit{even} character. }
	\label{table:table1}
\end{table*}

\FloatBarrier

\end{document}